\documentclass[aps,pre,twocolumn,groupedaddress]{revtex4-2}
\usepackage{amsmath}
\usepackage{amsfonts}
\usepackage{graphicx}
\usepackage{csquotes}
\usepackage{color,soul}
\begin{document}

% Use the \preprint command to place your local institutional report
% number in the upper righthand corner of the title page in preprint mode.
% Multiple \preprint commands are allowed.
% Use the 'preprintnumbers' class option to override journal defaults
% to display numbers if necessary
%\preprint{}

%Title of paper
\title{Kinetic random-field nonreciprocal Ising model}
% repeat the \author .. \affiliation  etc. as needed
% \email, \thanks, \homepage, \altaffiliation all apply to the current
% author. Explanatory text should go in the []'s, actual e-mail
% address or url should go in the {}'s for \email and \homepage.
% Please use the appropriate macro foreach each type of information

% \affiliation command applies to all authors since the last
% \affiliation command. The \affiliation command should follow the
% other information
% \affiliation can be followed by \email, \homepage, \thanks as well.
\author{Arjun R}
\author{A. V. Anil Kumar}
\email[]{anil@niser.ac.in}
%\homepage[]{Your web page}
%\thanks{}
%\altaffiliation{}
\affiliation{School of Physical Sciences, National Institute of Science Education and Research, Jatni, Bhubaneswar 752050, India}
\affiliation {Homi Bhabha National Institute, Anushakti Nagar, Mumbai, India}
%Collaboration name if desired (requires use of superscriptaddress
%option in \documentclass). \noaffiliation is required (may also be
%used with the \author command).
%\collaboration can be followed by \email, \homepage, \thanks as well.
%\collaboration{}
%\noaffiliation
\begin{abstract}
We introduce and analyse the \emph{kinetic random-field nonreciprocal Ising model}, which combines bimodal (double-delta) diffusive disorder with pairwise nonreciprocal interactions between two species. By employing mean-field and effective-field theories alongside kinetic Monte Carlo simulations (3D Glauber dynamics), we identify a nonequilibrium tricritical (Bautin) point separating continuous Hopf-type transitions from discontinuous saddle-node-of-limit-cycle (SNLC) transitions. For weak disorder below a critical value, collective oscillations (the “swap” phase) emerge via a supercritical Hopf bifurcation; above this value, the transition becomes first-order (SNLC), exhibiting hysteresis and characteristic Binder-cumulant signatures. Finite-size scaling of the susceptibility confirms the distinct critical and discontinuous behaviours in the Hopf and SNLC regimes, respectively (yielding effective exponents $\approx1.96$ and $\approx3.0$). In the first-order regime, the swap phase requires a threshold nonreciprocity that increases with disorder strength. The above conclusions also hold qualitatively for a random field sampled from a double-Gaussian distribution. Finally, we identify a droplet-induced swap phase at higher disorder that cycles through eight metastable states, driven by droplet nucleation in the dynamical free-energy landscape. These findings reveal how disorder and nonreciprocity generate rich nonequilibrium criticality relevant to driven and active systems.
\end{abstract}
% insert suggested keywords - APS authors don't need to do this
%\keywords{}
%\maketitle must follow title, authors, abstract, and keywords
\maketitle
% body of paper here - Use proper section commands
% References should be done using the \cite, \ref, and \label commands
\section{Introduction}
Phase transitions and their universality are key concepts in Statistical Physics, especially for systems exhibiting collective behaviour \cite{landau, kadanoff, stanley, wilson}. Many of these systems function far from equilibrium, where the presence of nonzero probability currents and particular symmetries can give rise to non-equilibrium phase transitions. Such phase transitions occur across a wide range of systems, including driven and active systems \cite{cates,buttinoni,Grafke,geyer,martin}, biology \cite{liu}, neuroscience \cite{mazzucato,jercog}, fluid dynamics \cite{grafke1}, climate science \cite{simonnet}, population dynamics \cite{kwon,buhl}, etc. In many of these non-equilibrium systems, the effective interactions can be nonreciprocal---interactions that violate action–reaction symmetry between different particles or species. Recent studies show that nonreciprocal interactions produce new collective behaviours—such as time-dependent ordered phases and transitions involving exceptional points—that go beyond traditional free-energy minimisation \cite{fruchart,you,dadhichi,loos}. Experiments reveal that nonreciprocal interactions drive self-assembly in biological \cite{tan} and synthetic \cite{ibele} systems. In living organisms, these interactions often stem from information asymmetry due to environmental factors (e.g., airflow) or sensory mechanisms (e.g., vision) \cite{dadhichi}. An interesting observation is the emergence of polarity from self-organisation of self-propelled active molecules in a binary mixture of non-polar particles with nonreciprocal interactions \cite{soto}. Interactions among more than two species can produce complex behaviours like periodic clustering and explosive dynamics, as seen in enzymes cyclically linked in chemical configuration space \cite{reboul}. Over the past decade, research on nonreciprocal interactions has expanded our understanding of diverse collective behaviours in inherently nonequilibrium systems.

Recent models have explored novel features arising from nonreciprocal interactions. The nonreciprocal extension of the Cahn–Hilliard model provides a framework to understand phase separation in binary mixtures interacting nonreciprocally \cite{saha}. Nonreciprocal extensions of Ising \cite{seara,rajeev}, XY \cite{rouzaire} and Heisenberg \cite{bhatt} models have been studied using Monte Carlo simulations, and these studies have uncovered changes in the phase behaviour as well as critical exponents. A simple yet insightful minimal model of nonreciprocity is offered by recent extensions of the Ising model incorporating asymmetric couplings between two species of spins \cite{avni2025nonreciprocal}. Mean-field analysis of this nonreciprocal Ising model predicts not only the conventional disordered and statically ordered phases but also a time-dependent “swap” phase, characterised by out-of-phase oscillations between the two species. Numerical studies show fluctuations and spatial structure enrich this behaviour: droplet excitations disrupt static order, spiral defects destabilise the swap phase in low dimensions, and in three dimensions, nonreciprocity can alter universality classes and stabilise spatiotemporal oscillations \cite{avni2025nonreciprocal}. These results motivate further study of how random-field disorder impacts nonreciprocal critical phenomena.

Disorder that couples directly to the order parameter — as realised in the paradigmatic random-field Ising model (RFIM) — profoundly modifies critical behaviour and, in low enough spatial dimension, can destroy long-range order.  The seminal Imry–Ma argument and subsequent scaling analyses established that arbitrarily weak random fields destabilise ordered states in sufficiently low dimensions (for the Ising case, one finds $d_{\rm lc}=2$) and motivated a general scaling framework for RFIM criticality \cite{imry1975random}. Over the last decade, high-precision zero-temperature studies have clarified the equilibrium picture in physically relevant dimensions.  In three dimensions, extensive $T\!=\!0$ ground-state simulations and refined finite-size-scaling analysis have converged on a single universality class for the paramagnetic–ferromagnetic transition once strong scaling corrections are properly accounted for, resolving earlier contradictory claims of nonuniversal or first-order behaviour \cite{fytas2013universality}. In four and five dimensions, analogous numerical studies, together with refined finite-size analysis, have further sharpened the determination of critical exponents and the modified hyperscaling relations appropriate for a zero-temperature fixed point \cite{fytas2023finite}. A  key theoretical development in this context is the Parisi–Sourlas observation that, under an emergent supersymmetry (SUSY) of the disorder-averaged field theory, the RFIM in $D$ dimensions would display {\it dimensional reduction} to the pure (no-disorder) Ising model in $D-2$ \cite{bray1985scaling}.  However, this elegant perturbative prediction can be spoiled by nonperturbative effects (avalanches, droplets, multiple metastable solutions), which break the SUSY and invalidate dimensional reduction in sufficiently low dimensions.  Extensive analytical and numerical work has clarified the regimes in which SUSY survives and where it fails; comprehensive summaries can be found in
references \cite{fytas2018review, rychkov2303four, tissier2011supersymmetry, barber2001monte, fytas2019evidence, villain1984nonequilibrium}. Finally, and most relevant to the present work, equilibrium (quenched) RFIM physics does not necessarily extend to systems with mobile or diffusive disorder.  Kinetic / competing-kinetics models in which the local field distribution is updated on comparable time scales to the spins---a natural model of diffusive or rapidly fluctuating disorder---generically produce nonequilibrium steady states \cite{alonso1992non, garrido1994kinetic, marro2005nonequilibrium}. Mean-field treatments and Monte Carlo simulations of such kinetic random-field models reveal qualitatively new phenomena, including disorder-induced tricriticality and first-order transitions—features completely absent in the equilibrium quenched RFIM \cite{crokidakis2010nonequilibrium}. These systems, often termed the nonequilibrium random-field Ising model (NRFIM), thus constitute a distinct and intrinsically nonequilibrium universality class. Moreover, effective-field calculations by Costabile \emph{et al.} \cite{costabile2012phase} align with these numerical observations. Incorporating nonreciprocal interactions into this framework provides a natural and intriguing next step, opening the possibility that the interplay of dynamical disorder and nonreciprocal couplings may reshape the phase diagram in fundamental ways, potentially generating entirely new regimes of collective behaviour.

In this work, we introduce the random-field nonreciprocal Ising model, which combines bimodal random fields (diffusive) with antisymmetric inter-species coupling. Using mean-field theory (MFT), effective-field theory (EFT) that incorporates the self-spin correlations, and three-dimensional kinetic Monte Carlo simulations, we characterise the transitions among disordered, time-dependent swap, and droplet-induced swap phases.   Our key findings are (i) identifying a nonequilibrium tricritical point (Bautin point) that separates continuous Hopf-type onsets of collective oscillations from discontinuous saddle-node-of-limit-cycle (SNLC) transitions; (ii) a monotonic increase in the nonreciprocity threshold with random field strength in the discontinuous regime; and (iii) identifying a new droplet-induced swap phase cycling through eight metastable states at high random field and subcritical nonreciprocity. These results are supported by finite-size scaling, probability distribution of the order parameter, and Binder cumulant analysis. 

\section {Model and methods}
The RFIM is a fundamental model for disordered systems, described by the Hamiltonian:
\begin{equation}
{\cal H} = -J\sum_{\langle i,j\rangle}\sigma_i \sigma_j - \sum_i h_i\sigma_i,
\end{equation}
where $\sigma_i$ represents the Ising spins, $J$ is the nearest-neighbor ferromagnetic interaction, and $h_i$ represents the local random fields. Motivated by the fully antisymmetric nonreciprocal Ising model proposed by Avni et al. \cite{avni2025nonreciprocal}, we extend this to a two-species nonreciprocal Ising model with bimodal disorder. Since nonreciprocal interactions preclude the definition of a global Hamiltonian, the dynamics are governed by the local selfish energy of a spin $\sigma_i^\alpha$, defined as:
\begin{equation}
    E^\alpha_i = -J\sum_{j_{nn}}\sigma^\alpha_i\sigma^\alpha_{j_{nn}} - K_{\alpha\beta}\sigma^\alpha_i\sigma^\beta_i-h_i\sigma^\alpha_i,    
\end{equation}
where, $J>0$ denotes the reciprocal intra-species coupling, while $K_{AB} = -K_{BA} = K > 0$ represents the antisymmetric inter-species coupling. This antisymmetry introduces explicit non-reciprocity, breaking the action-reaction symmetry between species A and B. Here $\sum_{j_{nn}}$ runs over nearest neighbours of site $i$, and $h_i$ is a site-dependent random field drawn from a bimodal (double-delta) distribution.
\begin{equation}\label{\theequation}
    P(h_i) = \frac{1}{2}[\delta(h_i-h) + \delta(h_i +h)]
\end{equation}
with $h>0$ the field strength. A bimodal (double-delta) distribution is chosen because it is the simplest probability distribution which produces a tricritical point in the NRFIM. We consider the dynamics to proceed by single-spin flip Glauber kinetics corresponding to the transition rate
\begin{equation}
    \omega(\sigma_i^\alpha \to -\sigma_i^\alpha) = \frac{1}{2\tau}[1-\tanh(\Delta E_i^\alpha/2k_BT)],
\end{equation}
where $\tau$ is the correlation time, $\Delta E_i^\alpha$ is the change in selfish energy due to a single spin flip, and $T$ is the temperature. The dynamics of magnetisation of the different species are described by the master equation
\begin{equation}\label{\theequation}
     \frac{\partial P(\sigma, t)}{\partial t} = \sum_{\sigma'} [P(\sigma, t)\omega(\sigma|\sigma') - P(\sigma', t)\omega(\sigma'|\sigma)],   
\end{equation}
where $P(\sigma, t)$ is the probability of finding the configuration $\sigma$ at time $t$ and $\omega(\sigma|\sigma')$ is the transition rate from the state $\sigma$ to $\sigma'$.

\subsection {Mean and effective field theory}
 To obtain the time-evolution equations for the magnetisation of each species, we analyse the master equation \eqref{5} within both the mean-field theory framework and the effective-field theoretical approach.
Multiplying both sides of \eqref{5} by $\sigma_i^\alpha$ and taking the average, we get
\begin{align}
    \tau\frac{dm_i^\alpha}{dt} &= -m_i^\alpha \nonumber\\
    &\quad+ \left\langle\tanh\left[\frac{1}{k_BT}(J\sum_{j_{nn}}\sigma_{j_{nn}}^\alpha + K_{\alpha\beta}\sigma_i^\beta + h_i)\right]\right\rangle.
\end{align}
The mean field approximation involves using $\langle f(\sigma)\rangle = f(\langle\sigma\rangle)$. Furthermore, taking the configurational average, considering the magnetisation to be homogeneous and rescaling time and space \cite{avni2025dynamical}, we arrive at the MFT result
\begin{align}\label{\theequation}
        \frac{d M_\alpha(t)}{d t} &= -M_\alpha + \frac{1}{2}[\tanh(\tilde{J}M_\alpha + \tilde{K}_{\alpha\beta}M_\beta + \tilde{h}) \nonumber\\
        &\quad + \tanh(\tilde{J}M_\alpha + \tilde{K}_{\alpha\beta}M_\beta - \tilde{h})],
\end{align}
where $\tilde{J} = 2dJ/k_B T$ (d is the spatial dimension). $\tilde{K}_{\alpha\beta} = K_{\alpha\beta}/k_B T$, and $\tilde{h} = h/k_B T$. For simplicity, we refer to the scaled variables using their original names. This result is found for the quenched random field; nevertheless, Alonso $\And$ Marro \cite{alonso1992non} show that, at zeroth-order mean-field and for the transition rate \eqref{4}, the non-equilibrium random-field model (diffusive) yields the same result as the quenched mean-field approximation. One can also refer to \cite{marro2005nonequilibrium} for further discussion.

To include self-spin correlations, we introduce the differential operator technique into $\tanh(\tilde{J}\sum^z_\delta\sigma^\alpha_{i+\delta} + \tilde{K}_{\alpha\beta}\sigma_i^\beta + \tilde{h}_i)$ as per the EFT technique \cite{costabile2012phase}. This is done before taking the expectation value inside the $\tanh$ function.
\begin{align}
    &\quad\tanh(\tilde{J}\sum^z_\delta\sigma^\alpha_{i+\delta} + \tilde{K}_{\alpha\beta}\sigma_i^\beta + \tilde{h}_i) = \nonumber\\ 
     &\exp{(\tilde{J}\sum^z_\delta\sigma^\alpha_{i+\delta} +  \tilde{K}_{\alpha\beta}\sigma_i^\beta)} \tanh(x+\tilde{h}_i)|_{x=0},
\end{align}
$z$ is the coordination number. This leads us to, after some approximations, the EFT result
\begin{equation}\label{\theequation}
    \frac{d M_\alpha(t)}{\partial t} = -M_\alpha +  \sum_{p=0}^6A_p^\alpha(M_\alpha)^p.
\end{equation}
Here, 
\begin{align}\label{\theequation}
    A_p^\alpha &= \frac{6!}{p!(6-p)!}\cosh^{6-p}{(\tilde{J}D_x)}\sinh^{p}{(\tilde{J}D_x)}\nonumber\\&\quad \times[\cosh(\tilde{K}_{\alpha\beta}D_x) + M_\beta\sinh(\tilde{K}_{\alpha\beta} D_x)]F(x)|_{x=0},
\end{align}
where $F(x) = (1/2)[\tanh(x+\tilde{h}) + \tanh(x-\tilde{h})]$ and $D_x = d/dx$. The coefficients $A_p^\alpha$ can be calculated using the property $\exp{(aD_x)}g(x) = g(x+a)$ (see Appendix~\ref{Appendix_A}). In EFT, correlations between different spins are neglected, and the same-spin correlations are retained  \cite{costabile2012phase} (in MFT, all the spin correlations are discarded). Thus, EFT gives a better approximation than MFT.

The steady and time-dependent order is characterised via the synchronisation amplitude $R$, and an angular momentum like variable $S$ \cite{avni2025nonreciprocal},
\begin{align}\label{\theequation}
R &= \sqrt{\frac{M_A^2+M_B^2}{2}} \quad & S &= M_A\frac{\partial M_B}{\partial t} - M_B\frac{\partial M_A}{\partial t}.
\end{align}
$R$ and $S$ quantify the alignment of spins and rotation of the vector ($M_A$, $M_B$), respectively.

\subsection{Monte Carlo simulations}
We perform 3D Monte Carlo simulations (since the swap phase is stable in 3D \cite{avni2025nonreciprocal}) with single-spin flip Glauber updates. The lattice is chosen to be simple cubic with periodic boundary conditions. The time unit is taken as one Monte Carlo sweep. Random field $h_i$, sampled from \eqref{3}, is updated every unit of time for competing kinetics \cite{crokidakis2010nonequilibrium}. In practice, we discard an initial equilibration period of $10^5$ Monte Carlo sweeps (greater than the time required for the order parameters $R$ and $S$ to stabilise). The measured quantities used in the finite-size scaling analysis are averaged over $10$ independent realisations.  Sample-to-sample fluctuations were quantified via a nonparametric bootstrap over the realisations (we used $N_{\mathrm{boot}}=1000$ resamples).

The two-dimensional histogram \(P(R, E)\), where $E$ denotes the  average selfish energy, is constructed from the post-equilibration samples.  Marginalising over $E$ yields the distribution $P(R)$ (which is equivalent to  computing  $P(R)$ from the time series of $R(t)$) with $R =\sqrt{(M_A^2+M_B^2)/2}$ and each $P(R)$ is normalised to unit area. For visualization, we apply Gaussian smoothing to reduce sampling noise, overlay the smoothed $P(R)$ curves for different $(\tilde J, \tilde h)$, and locate peaks via a prominence criterion to identify and track bimodality.
\section{Results and discussion}
\subsection{Mean-field and effective-field predictions}
\begin{figure}
 \includegraphics[width=0.95\linewidth]{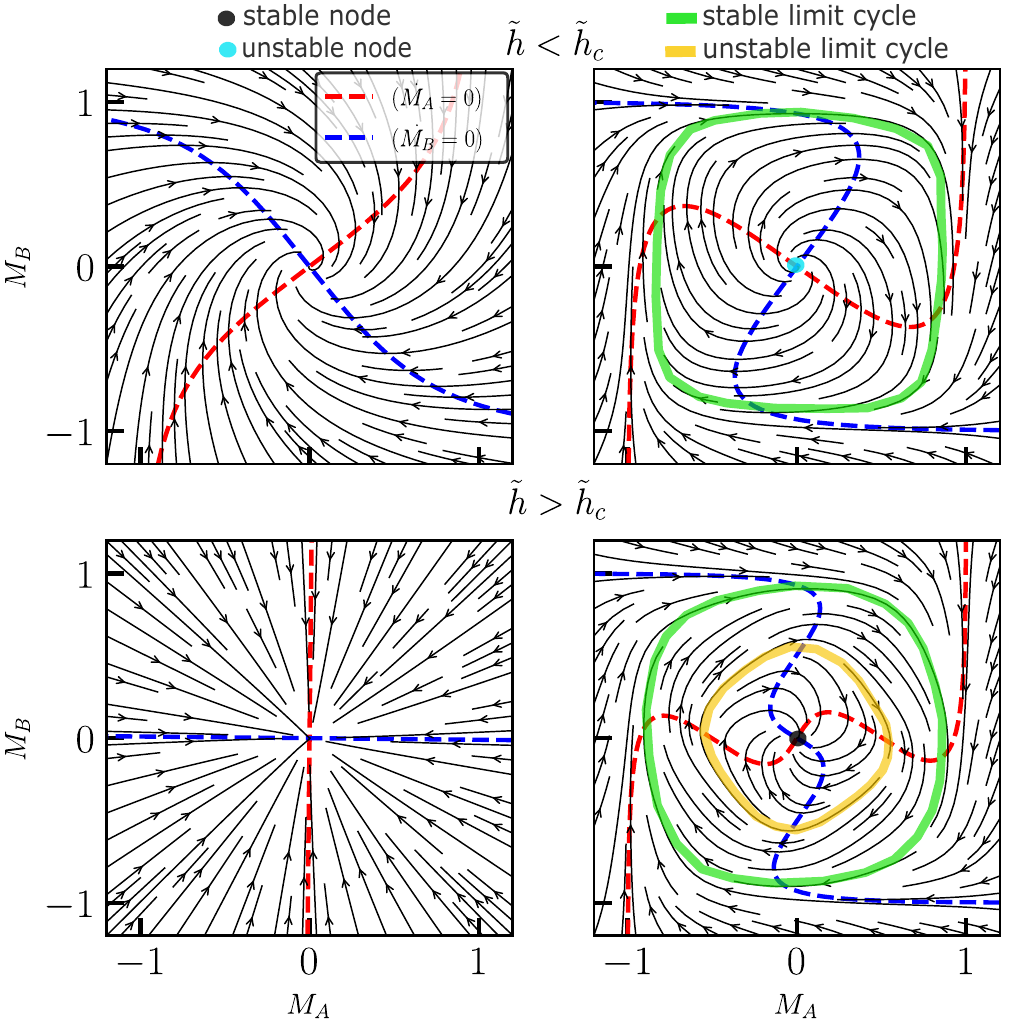}
\caption{Representative phase portraits illustrating two regimes: top row, $\tilde h<\tilde h_c$ (parameters: $\tilde J=0.5$ for the disordered fixed point and $\tilde J=2.0$ for the swap state, with $\tilde K=1$ and $\tilde h=0.5$); bottom row, $\tilde h>\tilde h_c$ (parameters: $\tilde J=0.5$ disordered and $\tilde J=3.0$ swap, with $\tilde K=1$ and $\tilde h=1.5$). These panels show that the system crosses from a supercritical Hopf to an SNLC bifurcation as disorder increases past $\tilde h_c$. These results are obtained from the mean-field theory. \label{1}}
\end{figure}

\begin{figure}
 \includegraphics[width=0.95\linewidth]{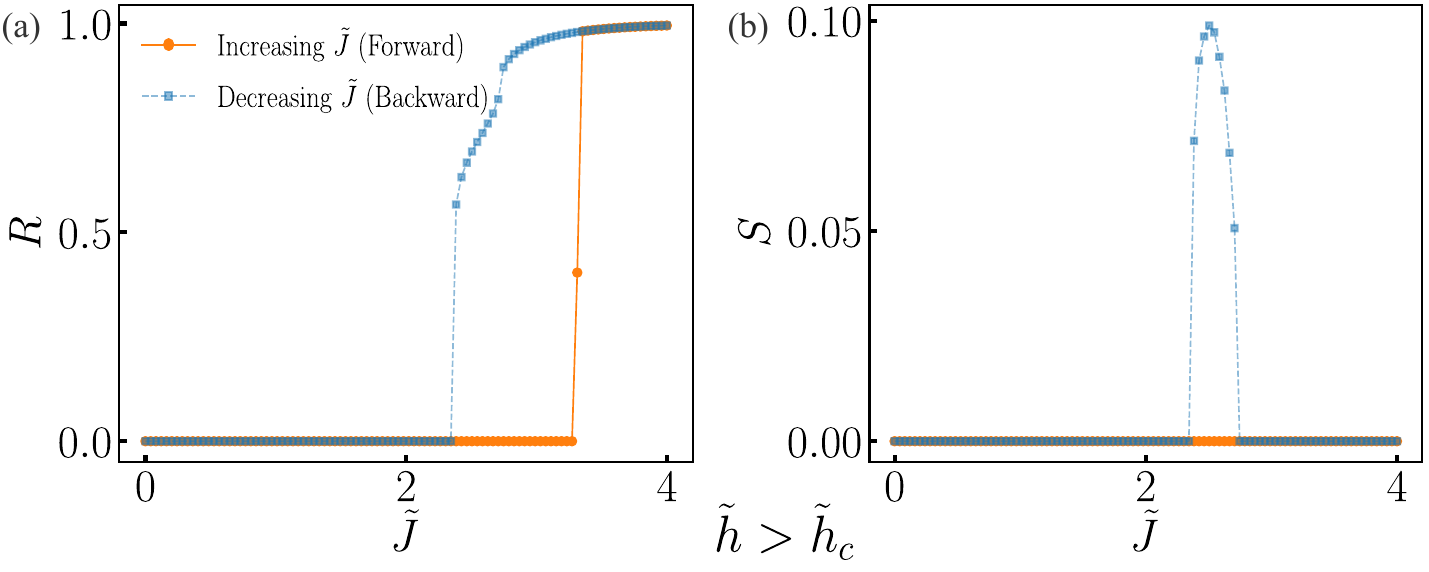}
\caption{Hysteresis of the (a) synchronisation amplitude $R$ and (b) rotation measure $S$ for $\tilde h>\tilde h_c$  from mean-field theory calculations. Here we fix $\tilde h = 1.2$ and $\tilde K = 0.3$, for which we observe a clear signature of a discontinuous transition; EFT gives similar trends. \label{2}}
\end{figure}

A detailed numerical investigation of Eq.~\eqref{7} uncovers a rich bifurcation landscape whose structure depends sensitively on the magnitude of the random field. For weak disorder, the fixed point at the origin—which represents the disordered phase—loses stability through a supercritical Hopf bifurcation as the effective coupling $\tilde J$ is increased. Beyond this point, the dynamics settle into a stable limit cycle, signalling the onset of persistent oscillations. This oscillatory regime, termed the swap phase, arises via a continuous dynamical phase transition \cite{avni2025nonreciprocal}, as shown in the top panels of Fig.~\ref{1}. When the disorder strength is increased, however, this scenario is qualitatively altered. The emergence of oscillations is no longer smooth: instead, the system undergoes an SNLC bifurcation, illustrated in the bottom panels of Fig.~\ref{1}. In this case, a stable and an unstable limit cycle are simultaneously created, and the latter undergoes a subcritical Hopf bifurcation with the stable spiral at the centre leading to a discontinuous transition into the swap phase. 
The crossover between the continuous (Hopf) and discontinuous (SNLC) regimes occurs at a critical disorder strength $\tilde h_c \approx 0.658$. This value is obtained by expanding the right-hand side of Eq. \eqref{7} and identifying the point at which the cubic coefficient in the normal form changes sign. The resulting change in bifurcation character signifies the presence of a nonequilibrium tricritical point—specifically, a Bautin (generalised Hopf) bifurcation \cite{kuznetsov1998elements}. To our knowledge, the only previous lattice model in which a Bautin bifurcation has been identified is the dissipative Curie–Weiss system with bimodal random fields \cite{collet2019effects}. Additional evidence for the first-order nature of the transition in the $\tilde h > \tilde h_c$ regime is provided by the appearance of hysteresis in the order parameters $R$ and $S$, shown in Fig.~\ref{2}. This hysteresis loop reflects bistability near the transition and is a clear hallmark of discontinuity.

\begin{figure}[h!]
 \includegraphics[width=0.95\linewidth]{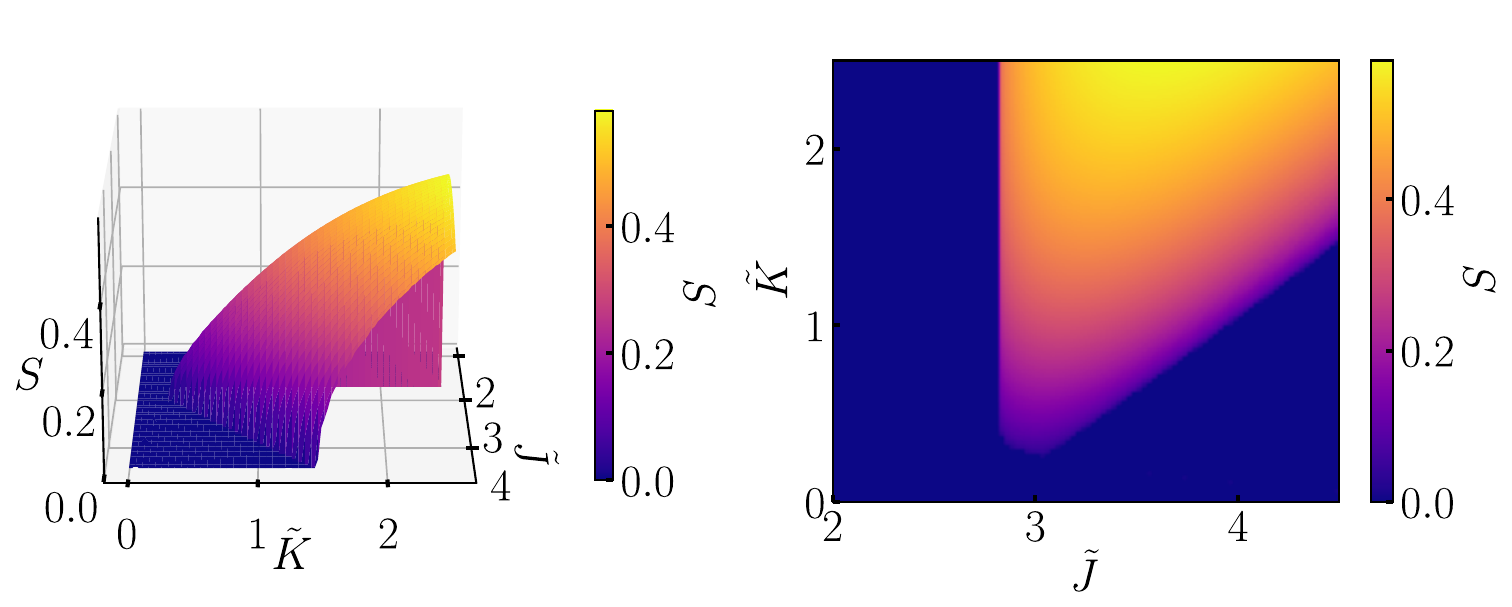}
\caption{Dependence of the swap phase on nonreciprocity $\tilde K$ and interaction strength $\tilde J$: when $\tilde h>\tilde h_c$, a finite threshold $\tilde K_c$ is required to sustain oscillations, and the swap phase vanishes for $\tilde K<\tilde K_c$ (illustrated for $\tilde h=1.5$). The results shown here are obtained within the mean-field approximation. \label{3}}
\end{figure}

To refine the predictions beyond MFT, we also employed an EFT approach \cite{costabile2012phase}, which includes self–spin correlations via the differential-operator technique. While EFT preserves the same bifurcation topology as MFT---namely, the appearance of Hopf and SNLC bifurcations and a Bautin point---it shifts the critical lines quantitatively. In particular, EFT predicts the Bautin point to occur at significantly larger disorder strengths, in the range $1.35 < \tilde h < 1.40$, compared to the MFT value of $\tilde h_c \approx 0.658$. Representative MFT-predicted phase boundaries at $\tilde h = 1.5$ are shown in Fig.~\ref{3}. From these illustrations, it is understood that for $\tilde h > \tilde h_c$, the swap phase does not exist below a critical threshold of nonreciprocity $\tilde K_c$. This highlights the crucial role of nonreciprocal coupling in sustaining time-dependent order in the presence of strong disorder.

\subsection{Monte Carlo results}

\begin{figure}[h!]
 \includegraphics[width=0.95\linewidth]{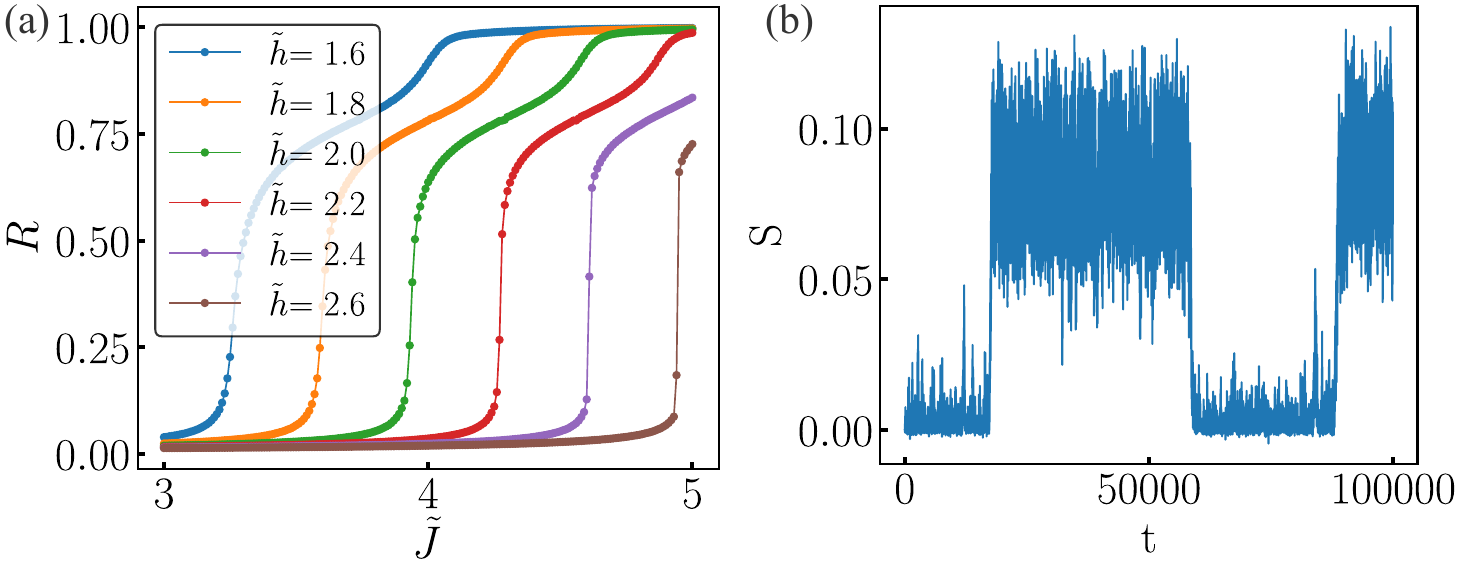}
\caption{Plotted at $\tilde K=0.35$. (a) Order parameter $R$ vs $\tilde J$ at $L=18$: as the random-field strength $\tilde h$ is increased, the onset of the swap phase sharpens, indicating a change from a continuous to a discontinuous transition. (b) Example time-series of the rotation measure $S$ in the coexistence region demonstrating abrupt switches between the disordered and swap phases. $t$ is measured in units of Monte Carlo sweeps (plotted at the first-order phase transition point and $\tilde h=2.4$). \label{4}}
\end{figure}

\begin{figure}[h!]
 \includegraphics[width=0.95\linewidth]{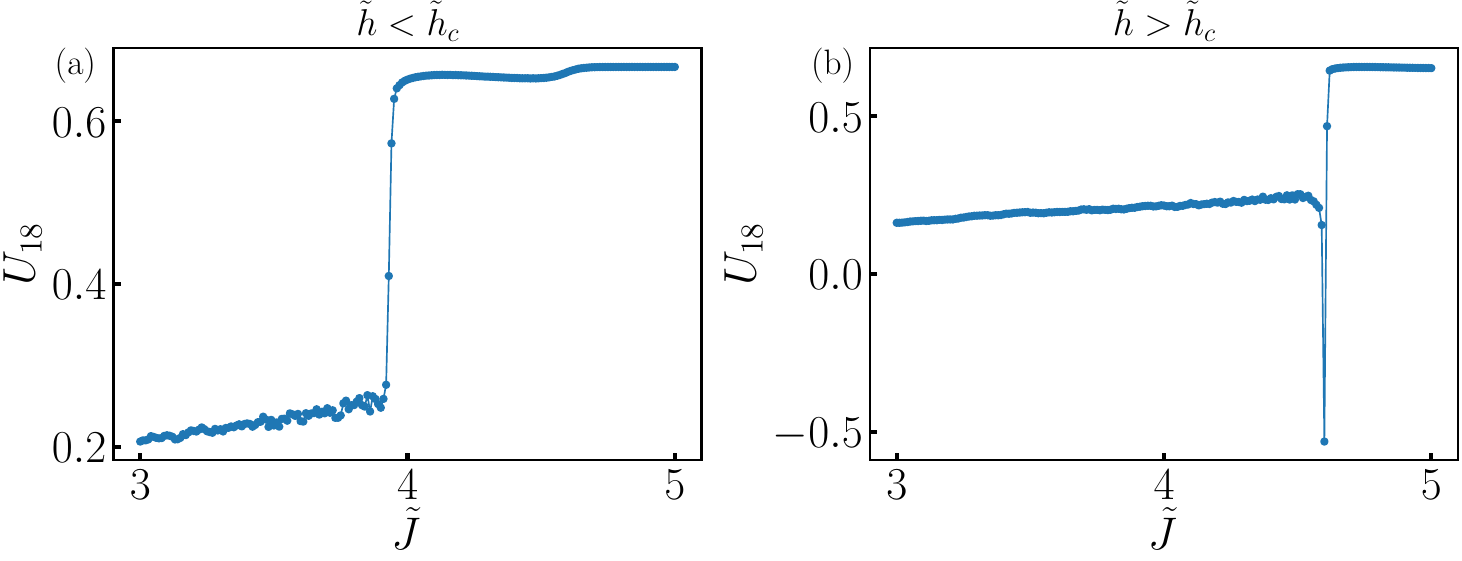}
\caption{Plotted at $\tilde K=0.35$. (a) Binder cumulant $U_{18}$ at $\tilde h = 2.0$ below the estimated $\tilde h_c$. (b) Binder cumulant $U_{18}$ at $\tilde h =2.4$ above the estimated $\tilde h_c$: a pronounced dip appears at the point of first-order phase transition. \label{5}}
\end{figure}

\begin{figure}[h!]
 \includegraphics[width=0.95\linewidth]{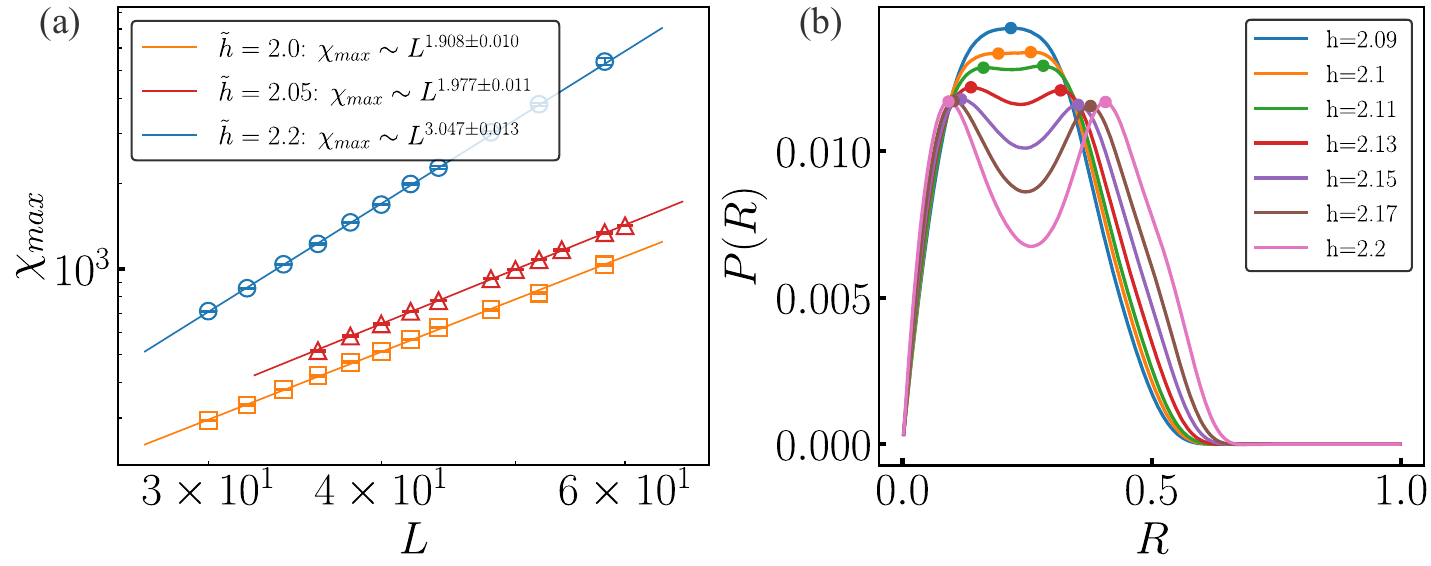}
\caption{(a) Finite-size scaling of the peak susceptibility $\chi_{\max}$ for $\tilde h=2.0,\,2.05,\,2.2$, $\tilde K = 0.35$ and using system sizes $L=30-60$. The scaling exponent crosses over from $\gamma/\nu\!<\!d$ to $\approx d$, consistent with a tricritical point separating a second-order (supercritical Hopf) and first-order (SNLC) behaviour. The simulation was performed for a bimodal (double-delta) random field distribution. (b) The probability distribution of the order parameter $R$ at the first-order phase transition point is plotted for different random field strengths at $L = 32$. \label{6}}
\end{figure}

We performed kinetic Monte Carlo simulations to validate the theoretical predictions obtained from both MFT and EFT. In these simulations, the system was allowed to evolve under three-dimensional Glauber dynamics, and the order parameters $R$ and $S$---which characterise collective time-dependent behaviour---were measured after a sufficient equilibration period. To identify the nature of the phase transitions, we analysed the synchronisation order parameter $R$ as a function of $\tilde J$, computed Binder cumulants, and performed finite-size scaling of the susceptibility at $\tilde K = 0.35$.  For sufficiently weak disorder $\tilde h$, the system exhibits a continuous transition from the disordered phase to the swap phase. As the disorder strength $\tilde h$ increases, the order parameter $R$ displays a sharp, discontinuous jump for $\tilde h > \tilde h_c$, indicating a first-order transition \cite{crokidakis2010nonequilibrium} (see Fig.~\ref{4}(a)).
Additionally, the Binder cumulant \cite{binder1984finite}, $U_L = 1 - \langle R^4\rangle/3\langle R^2\rangle^2$ behaves differently in the two regimes. For $\tilde h < \tilde h_c$ (tentative $\tilde h_c$ from Fig.~\ref{4}(a)), $U_L$ varies smoothly across the transition, consistent with a continuous onset of ordering (Fig.~\ref{5}(a)). However, for $\tilde h > \tilde h_c$, the cumulant develops a clear dip at the transition point (Fig.~\ref{5}(b)), a standard signature of a discontinuous phase transition \cite{tsai1998fourth}. 
We have also plotted $S(t)$ vs t (see Fig.~\ref{4}(b)) at the coexistence region, which shows the discontinuous jumps between disorder and swap phase. These numerical observations support the theoretical prediction of a nonequilibrium tricritical (corresponding to the Bautin point in MFT and EFT) point separating continuous and discontinuous transitions.
To further confirm the nature of the transition, we performed a finite-size scaling analysis of the peak susceptibility, $\chi_{\max}$ ($\chi = L^d(\langle R^2\rangle - \langle R\rangle^2)$, 
where $L$ is the linear system size and $d=3$ is the spatial dimensionality). The scaling behaviour of $\chi_{\max}$ provides a clear distinction between continuous and first-order transitions \cite{fisher1982scaling}. For $\tilde h = 2.0$ and $\tilde h = 2.05$, the susceptibility scales with a slope $1.908 \pm 0.010$ and $1.977 \pm 0.011$ in a log-log plot, respectively, consistent with continuous critical behaviour seen in \cite{avni2025nonreciprocal}, where $\chi_{\max} \sim L^{\gamma/\nu}$ with $\gamma/\nu < d$ and belongs to the 3D XY universality class.
In contrast, for stronger disorder values $\tilde h = 2.2$, we find the slope to be $3.047 \pm 0.013$, which is nearly equal to $d$. This scaling is characteristic of a first-order transition in three dimensions, where fluctuations scale extensively as $\chi_{\max} \sim L^d$ \cite{fisher1982scaling} (see Fig.~\ref{6}(a)). Together with the Binder cumulant and time-series analyses, the finite-size scaling of susceptibility provides consistent and converging evidence for a change in the character of the transition with increasing random-field strength $\tilde h$. Based on this scaling behaviour, we estimate the tricritical region---the crossover between continuous and discontinuous transitions---to lie within the range 2.05 $< \tilde h_c <$ 2.2. The quantitative values of the critical thresholds differ between MFT, EFT, and simulations due to the varying levels of approximation and inclusion of correlations.

To more precisely locate the tricritical point, we examine the probability distribution of the order parameter, $P(R)$, across varying strengths of the random field $\tilde{h}$, as depicted in Fig.~\ref{6}(b). In the regime where  $\tilde h > \tilde h_c$, the distribution is expected to exhibit a clear bimodal structure, reflecting the coexistence of two distinct stable states associated with the first-order transition. Conversely, at the tricritical value 
$\tilde h = \tilde h_c$, $P(R)$ should manifest as a broadened single-peaked distribution, indicating the critical merging of these two states into a continuous transition. Our numerical simulations for a system of linear size $L$ = 32 reveal that this bimodal characteristic appears at $\tilde{h}$ = 2.1 (with increasing length, the bimodality in the order parameter distribution becomes sharper). Based on these observations and accounting for finite-size effects, we estimate that, in the thermodynamic limit, the tricritical point is confined to the narrow range $2.05 < \tilde h_c < 2.10$.

\begin{figure}[h!]
 \includegraphics[width=0.95\linewidth]{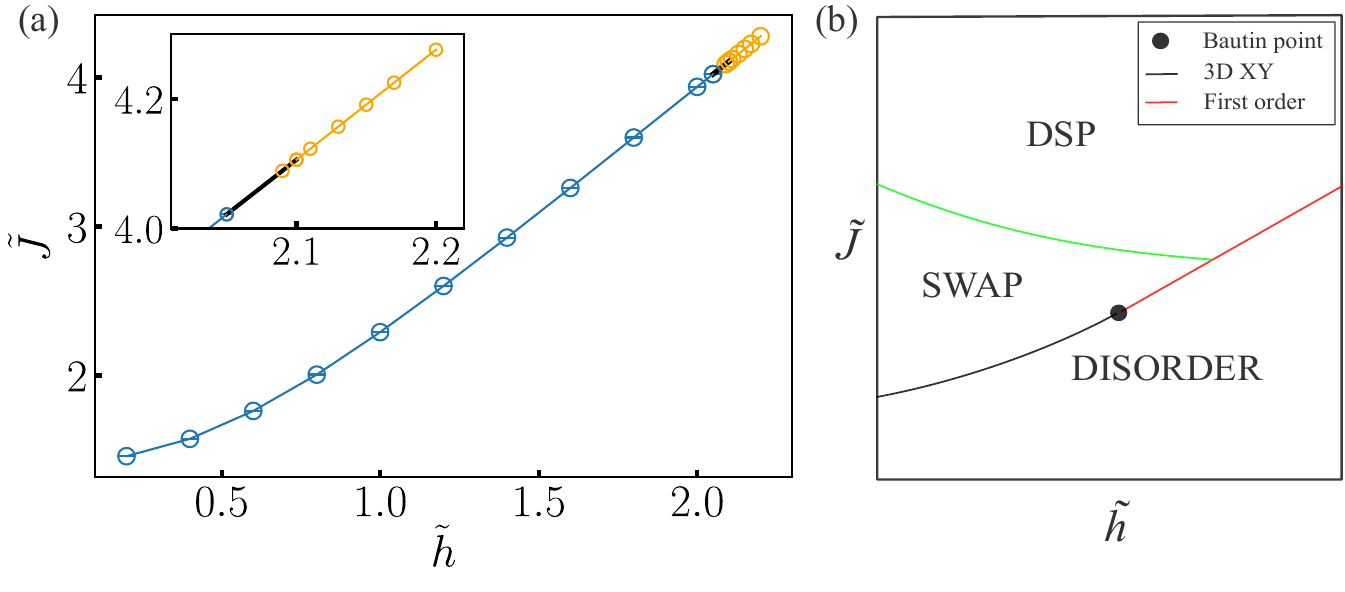}
\caption{(a) The black line marks the interval in which the nonequilibrium tricritical point is located, while the blue and yellow lines indicate continuous and discontinuous phase transitions, respectively. The diagram shows that the disorder–swap transition is continuous for $\tilde h < \tilde h_c$ and becomes first-order above $\tilde h_c$. The inset highlights the region near the tricritical point. Results are shown for $L = 32$ at $\tilde K = 0.35$, providing a qualitative representation of the phase diagram expected in the thermodynamic limit. (b) Schematic phase diagram of the random-field nonreciprocal Ising model, illustrating the phases and the phase transitions observed in the Monte Carlo simulations. DSP is the droplet-induced swap phase. \label{7}}
\end{figure}

Monte Carlo simulations indicate that the tricritical window lies within the range $2.05 < \tilde h_c < 2.10$, whereas the EFT yields a noticeably lower estimate, $1.35 < \tilde h_c < 1.40$. (with MFT predicting an even smaller value). This discrepancy highlights the substantial impact of fluctuations and the role of competing-kinetics dynamics in shifting the apparent location of the tricritical point. The results above are summarised in the phase diagram shown in Fig.~\ref{7}(a), plotted for $L$ = 32 and \(\tilde{K}=0.35\). The black line marks the interval in which the nonequilibrium tricritical point resides. We have also included in Fig.~\ref{7}(b),  a schematic phase diagram in the ($\tilde J$-$\tilde h$) plane for a fixed value of $\tilde K$, illustrating the different phases and the corresponding transitions. The black line denotes the line of continuous transitions separating the swap and disordered phase, and it terminates at a tricritical or Bautin point. Above the tricritical point, this transition becomes first-order; however, such a first-order disorder–to–swap transition occurs only when the nonreciprocity parameter
$\tilde K$  exceeds a critical threshold $\tilde K_c$.  The value of $\tilde K$ chosen for this schematic phase diagram is larger than $\tilde K_c$ in the vicinity of the Bautin point. As discussed in the next subsection, $\tilde K_c$ increases with $\tilde h$. Consequently, along the first-order line $\tilde K_c$ eventually surpasses the chosen $\tilde K$, at which point the swap phase disappears and is replaced by a droplet-induced swap phase. The transition from the disordered phase to this droplet-induced swap phase nevertheless remains first-order. The green line separates the swap and droplet-induced swap regimes. It should be noted that the existence and ultimate fate of the droplet-induced swap phase in three dimensions in the thermodynamic limit is still unresolved; see \cite{avni2025nonreciprocal} for further discussion.
 
\begin{figure}[h!]
 \includegraphics[width=0.95\linewidth]{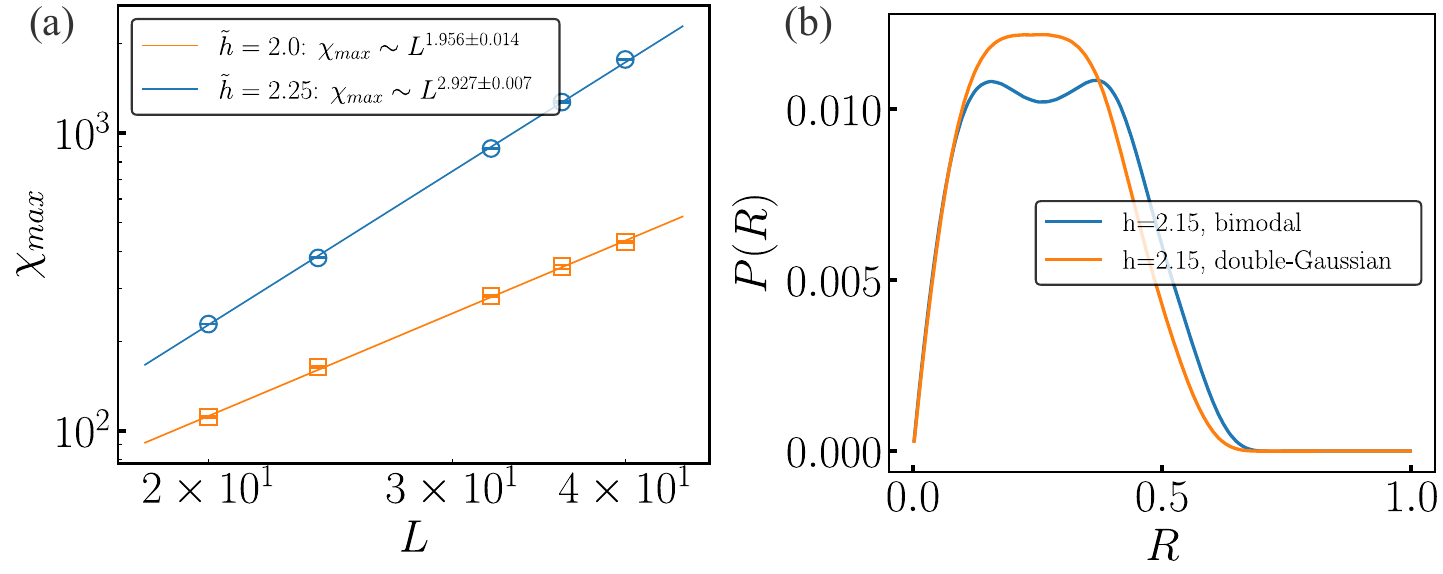}
\caption{(a) Finite-size scaling of the peak susceptibility $\chi_{\max}$ for $\tilde h=2.0,\,2.25$ using system sizes $L=20, 24, 32, 36, 40$. The scaling exponent crosses over from $\gamma/\nu\!<\!d$ to $\approx d$, consistent with a tricritical point separating a second-order (supercritical Hopf) and first-order (SNLC) behaviour. The simulation was performed for a double-Gaussian random field distribution. (b) Comparison between $P(R)$ of the different random-field probability distributions. Bimodal: double peak at $\tilde h = 2.15$. Double-Gaussian: single peak at $\tilde h = 2.15$. It is evident that the random field strength at the nonequilibrium tricritical point has shifted to a larger value in the latter. \label{8}}
\end{figure}

A double-delta random-field distribution is known to generate a large number of degenerate ground states in disordered spin systems \cite{hartmann1998ground}, raising the question of whether such degeneracy could artificially influence the nature of the phase transitions we observe. To assess the robustness of our conclusions against this possibility, we repeat our analysis using a smoother, continuous distribution—specifically, a double-Gaussian (equal-weight two-component Gaussian mixture) random field. In this alternative analysis, each site field  $h_i$
is independently drawn from an equal-weight two-component Gaussian mixture (double Gaussian),
\begin{equation}
    p(h_i)=\tfrac{1}{2}\,\mathcal{N}(\tilde h,\sigma^2)+\tfrac{1}{2}\,\mathcal{N}(-\tilde h,\sigma^2),
\end{equation}
\noindent which can be written explicitly as
\begin{align}
p(h_i) 
 &= \frac{1}{2\sqrt{2\pi}\,\sigma}\left[
      \exp\!\left(-\frac{(h_i-\tilde h)^2}{2\sigma^2}\right)
   \right. \\[4pt]
 &\quad\left.
      +\,\exp\!\left(-\frac{(h_i+\tilde h)^2}{2\sigma^2}\right)
   \right],
\end{align}
where $\sigma = 0.2$ denotes the standard deviation and $\tilde h$ is the random field strength. This distribution retains the symmetry properties of the double-delta case but eliminates the sharp discontinuities associated with discrete fields. Our finite-size scaling analysis indicates that the transition is continuous at $\tilde{h}=2.0$, whereas at $\tilde{h}=2.25$ it becomes first-order (see Fig.~\ref{8}(a), where the corresponding $\gamma/\nu$ values are indicated). A direct comparison of the probability distributions of the order-parameter $R$ for the bimodal and double-Gaussian fields at $\tilde{h} = 2.15$ (see Fig.~\ref{8}(b)) reveals that the tricritical point shifts to a higher random-field value when the double-Gaussian distribution is used. The key implication is that first-order behaviour persists even when the underlying disorder is continuous, which does not introduce the large ground-state degeneracies. This observation strongly supports the existence of a genuine tricritical point in the system, rather than an artefact arising from the discretisation of the random field. Nevertheless, we emphasise that finite-size effects may still influence the quantitative location of the tricritical point, and the precise value should therefore be interpreted carefully.

\subsection { Threshold nonreciprocity and tricritical point location}
Both theoretical analysis and Monte Carlo simulations consistently demonstrate that a minimum threshold of nonreciprocity, denoted by $\tilde K_c$, is necessary to sustain the time-periodic swap phase when the disorder strength exceeds the tricritical value ($\tilde h > \tilde h_c$). Across MFT and EFT, $\tilde h_c$ is independent of the nonreciprocity and we observe a monotonic relationship between $\tilde K_c$ and $\tilde h$ (Fig.~\ref{9}(a), see Appendix~\ref{Appendix_B} for details of the plotting procedure). On the other hand, in kinetic Monte Carlo simulations, we find that $\tilde h_c$ shifts to higher values with the growing nonreciprocity. This is illustrated in Fig.~\ref{9}(b); the bimodality of the order parameter $R$ decreases with increasing $\tilde K$, and eventually yields a single peak for which the selected $\tilde h$ is less than $\tilde h_c$.  
\begin{figure}[h!]
 \includegraphics[width=0.95\linewidth]{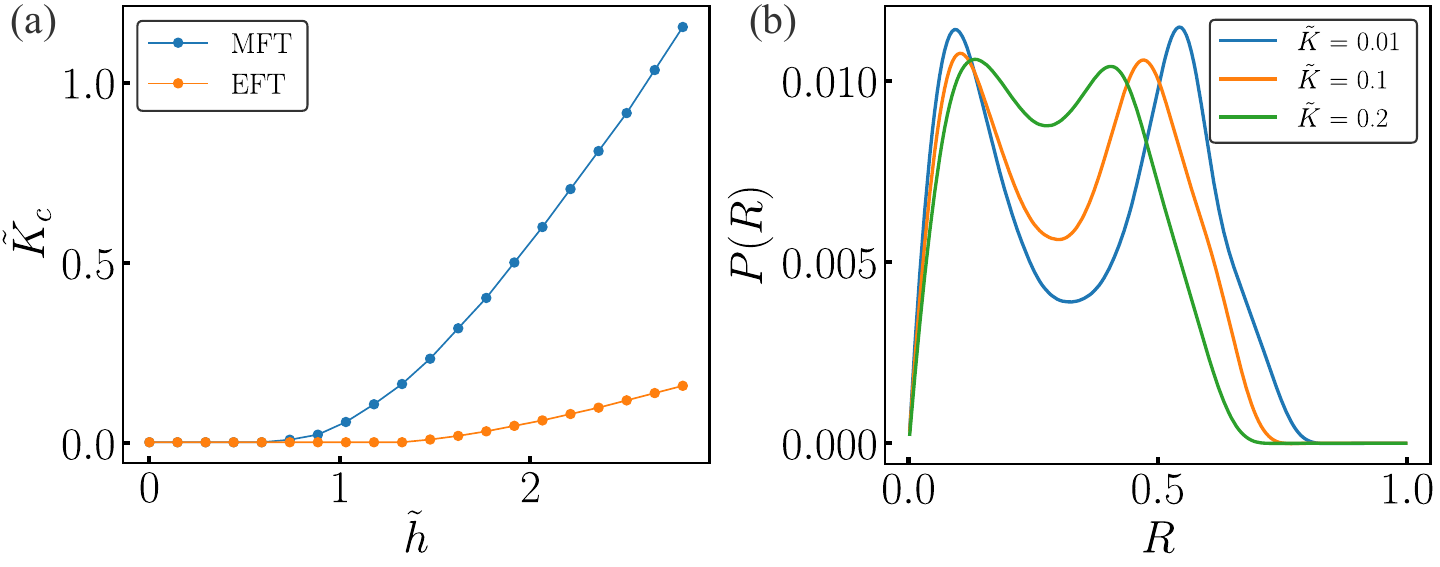}
\caption{(a) Analytical predictions for the threshold nonreciprocity $\tilde K_c$ as a function of random-field strength $\tilde h$, comparing MFT and EFT. EFT shifts the threshold relative to MFT due to the inclusion of self-spin correlations. (b) The probability distribution of the order parameter $R$ for different values of the nonreciprocity strength at $\tilde h = 2.05$ and $L = 24$. The bimodality of the distribution reduces with increasing $\tilde K$, suggesting that the random field strength at the nonequilibrium tricritical point is shifting to a higher value. \label{9}}
\end{figure}

 Monte Carlo simulations reveal that the critical disorder strength $\tilde h_c$ depends on the nonreciprocity parameter
 $\tilde{K}$. For disorder strengths exceeding $\tilde h_c$, sustaining the swap phase requires the nonreciprocity to exceed a critical threshold $\tilde K_c$. This critical threshold $\tilde K_c$ is found to be monotonously increasing with $\tilde h$. These observations reveal that
 $\tilde{h_c}$ can be written as a function of $\tilde{K}$, while $\tilde{K_c}$ is a function of $\tilde{h}$.  The nonreciprocity threshold curve $\tilde K_c(\tilde h)$, defined only for $\tilde h > \tilde h_c(0)$,  delineates the regime where the swap phase ceases to exist. In particular, the swap phase is absent when the conditions $\tilde h > \tilde h_c (\tilde K)$ and $\tilde K < \tilde K_c(\tilde h)$ are simultaneously satisfied.  These constraints partition the $\tilde{h} - \tilde{K}$ parameter space into three distinct regions, as defined by the tricritical curve $\tilde h_c(\tilde K)$ and the threshold nonreciprocity curve  $\tilde K_c(\tilde h)$. This is depicted in the schematic diagram shown in Fig.~11.  In region I,  the condition $\tilde h > \tilde h_c (\tilde K)$  is not  met, and the swap phase exists;  here the system undergoes a continuous transition from the disordered phase to the swap phase as $\tilde J$ increases, with no coexistence region, mirroring the behaviour of the standard nonreciprocal Ising model \cite{avni2025nonreciprocal}. In region II,  although $\tilde h > \tilde h_c (\tilde K)$, the condition $\tilde K < \tilde K_c(\tilde h)$ is not satisfied, allowing the swap phase to exist. Since this region is above the critical disorder strength, the transition from disordered state to swap phase with increasing $\tilde J$ is first order.  Finally, in region III, both the above conditions are satisfied, suppressing the swap phase  entirely; 
 instead, the system undergoes a first-order phase transition from the disordered to a droplet-induced swap phase.

\begin{figure}[h!]
 \includegraphics[width=0.95\linewidth]{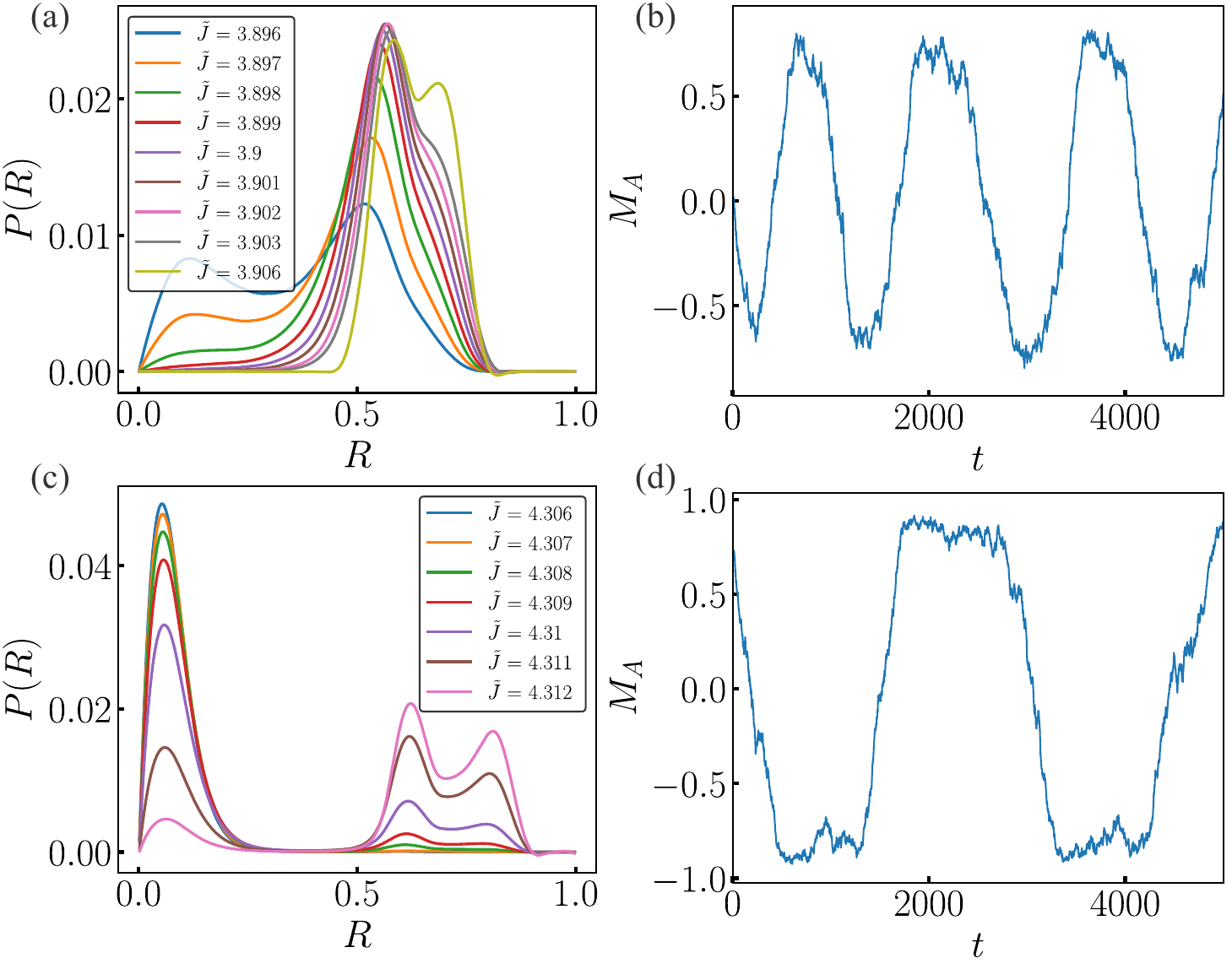}
\caption{(a) $P(R)$ at $\tilde K = 0.01$ and $\tilde h = 2.0$. At $\tilde J = 3.906$, the high $R$ mode becomes bimodal, indicating a transition from swap to droplet-induced swap phase. (b) $M_A$ vs $t$ at $\tilde J = 3.897$ for the case in (a). (c) $P(R)$ at $\tilde K = 0.01$ and $\tilde h = 2.25$. The high $R$ mode exhibits a bimodal structure from the moment it emerges, implying that the swap phase never existed. (d) $M_A$ vs $t$ at $\tilde J = 4.312$ for the case in (c). \label{10}}
\end{figure}

To show the existence of threshold nonreciprocity, we consider an example where the system has $\tilde K = 0.01$, and $\tilde h = 2.0$. The simulation results (Fig.~\ref{10}(a),(b)) indicate that the random field strength is well above the tricritical point since the probability distribution $P(R)$ is bimodal. We differentiate the swap phase from the droplet-induced swap phase by analysing the oscillation mechanism. While the swap phase exhibits homogeneous oscillations devoid of metastable states, the droplet-induced phase oscillates by jumping between metastable states via droplet formation (Fig.~\ref{12}(a)). This mechanism (or existence of metastable states) leads to a bimodal probability distribution, $P(R)$, at high $R$. Thus, in Monte Carlo simulations, a double peak in $P(R)$ at high R mode serves as the signature of the droplet-induced swap phase (refer to $\tilde J = 3.906$ in Fig.~\ref{10}(a)).
By inspecting the $P(R)$ for different $\tilde J$ values in Fig.~\ref{10}(a), we find that in our current example, the swap phase does exist. This means that $\tilde h = 2.0 > \tilde h_c(\tilde K = 0.01)$ and $\tilde K = 0.01 > \tilde K_c(\tilde h=2.0)$. Now we increase the random field strength to $\tilde h = 2.25$. We find from Fig.~\ref{10}(c),(d) that the high $R$ mode of $P(R)$ is always bimodal in the coexistence region, meaning that coexistence is between disorder and the droplet-induced swap phase and the swap phase never appeared in the system. Alternatively, $\tilde K = 0.01 < \tilde K_c(\tilde h = 2.25)$ at $\tilde h = 2.25>\tilde h_c(\tilde K = 0.01)$. These two scenarios suggest the presence of threshold nonreciprocity.

\begin{figure}[h!]
 \includegraphics[width=0.95\linewidth]{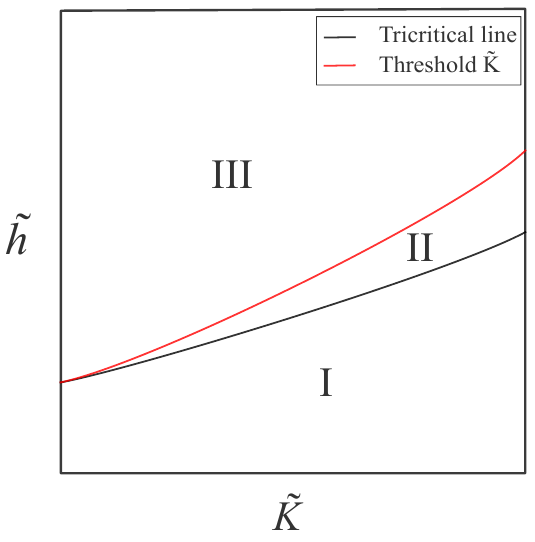}
\caption{Schematic diagram showing the nonequilibrium tricritical and threshold nonreciprocity (above which swap phase occurs in the $\tilde h > \tilde h_c$ regime) line. In Region III, the swap phase cannot exist. \label{11}}
\end{figure}

This trend indicates that stronger random fields tend to suppress homogeneous oscillatory behaviour unless offset by stronger nonreciprocal interactions. The appearance of threshold nonreciprocity can be understood physically: random fields induce locally preferred orientations that act as pinning centres, resisting coherent or homogeneous oscillations. As $\tilde h$ increases, stronger antisymmetric coupling (i.e., larger $\tilde K$) is needed to synchronise domains and overcome this local pinning. Fig.~\ref{11} shows the schematic diagram of the nonequilibrium tricritical line and the threshold nonreciprocity. In region III, the swap phase can not exist but it can exist in region I and II. The phase diagram in Fig.~\ref{7}(a) corresponds to the region I and II in Fig.\ref{11}.

\subsection { Droplet-induced swap dynamics and dynamical free-energy picture}
 In line with previous studies on nonreciprocal Ising systems \cite{avni2025nonreciprocal}, we find through Monte Carlo simulations that the statically ordered phase becomes unstable due to droplet excitations, particularly in strongly disordered regimes. Instead of settling into a time-independent ordered state, the system enters a droplet-induced swap phase, marked by spontaneous and recurring transitions between metastable configurations.
\begin{figure*}[t]
\centering
 \includegraphics[width=0.95\linewidth]{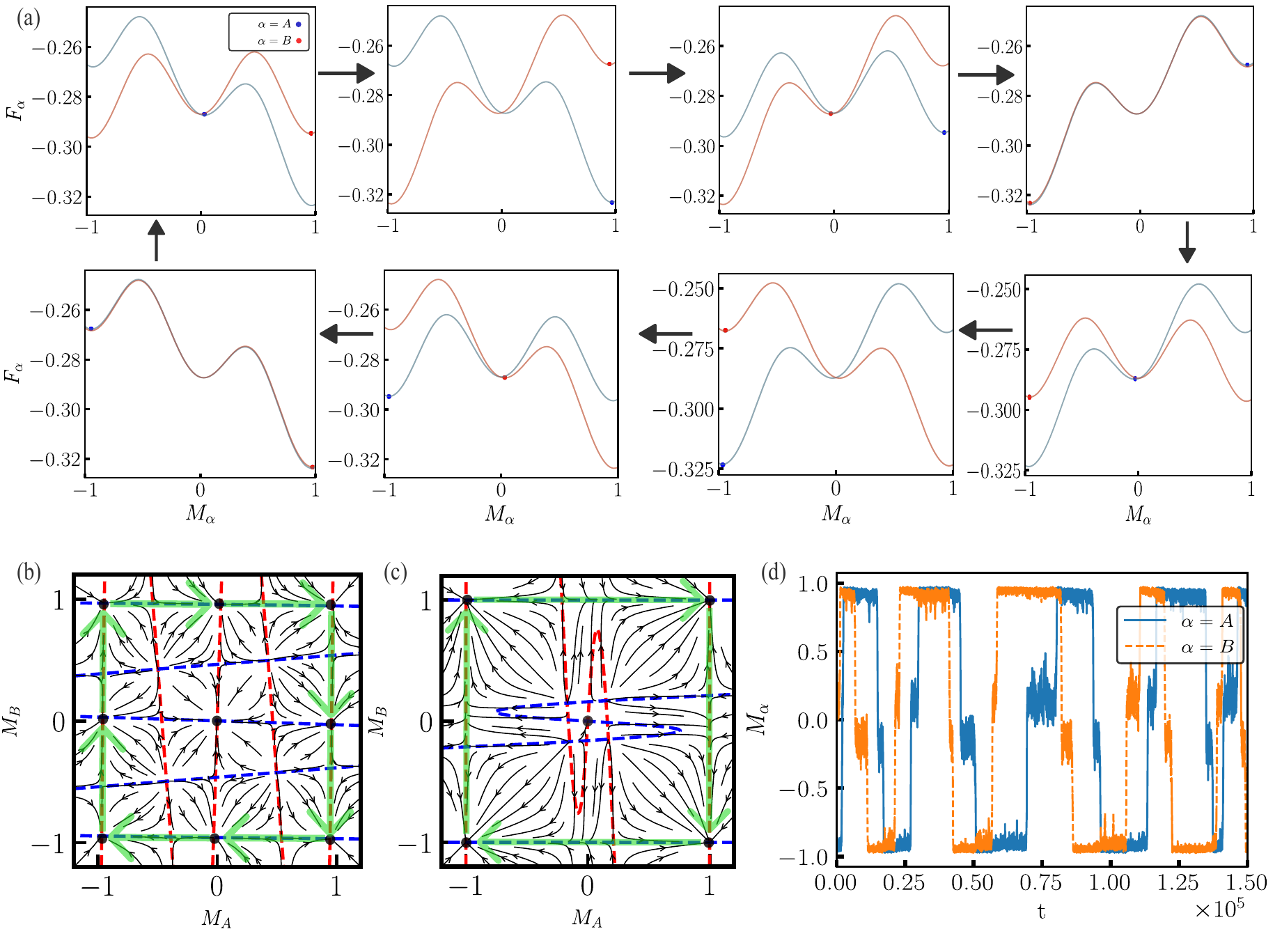}
\caption{(a) MFT: droplet-induced cycle visiting eight metastable states for $\tilde h>\tilde h_c$ (parameters: $\tilde J=3.3,\ \tilde K=0.1,\ \tilde h=1.6$). (b) Corresponding MFT phase portrait showing the cycle and its direction. Black dots represent the metastable fixed points (c) At larger coupling ($\tilde J=5.0$), the cycle reduces to four distinct states. (d) kinetic Monte Carlo results demonstrating droplet-induced swapping consistent with the MFT cycles (simulation: $\tilde J=5.04,\ \tilde K=0.0125,\ \tilde h=2.7,\ L=24$). \label{12}}
\end{figure*}
A new kind of behaviour emerges when an additional bimodal random field is introduced. If the random field strength exceeds the tricritical threshold ($\tilde h > \tilde h_c(\tilde K)$) and the nonreciprocity is below the critical value required to sustain coherent oscillations ($\tilde K < \tilde K_c(\tilde h)$), the system undergoes cyclic switching among a discrete set of eight metastable states, driven by local fluctuations and droplet nucleation events. A visualisation of this cyclic evolution is provided in the ancillary materials (as anc/Movie1.mp4), corresponding to simulations performed at $\tilde{J} = 5.038$, $\tilde{K} = 0.0125$, $\tilde{h} = 2.7$, and system size $L = 120$. For moderate values of the coupling $\tilde J$, the system cycles through eight distinct metastable states, as illustrated in Fig.~\ref{12}(a), (b). At higher values of $\tilde J$, the number of states involved in the cycle is reduced to four distinct configurations (as shown in Fig.\ref{12}(c)), which is the one observed in \cite{avni2025dynamical}. We interpret these findings in the framework of the dynamical equations governing the two spin species \cite{avni2025dynamical, kardar2007statistical, cates2019active, tauber2014critical}
\begin{equation}\label{\theequation}
    \frac{\partial M_\alpha(\textbf{r}, t)}{\partial t} = -\mu\frac{\delta F_\alpha[M_A, M_B]}{\delta M_\alpha(\textbf{r}, t)} + \eta_\alpha(\textbf{r}, t),
\end{equation}
where, $\mu$ is the generalized mobility and $\eta_\alpha$ is the white noise field such that it satisfies $\langle \eta_\alpha(\textbf{r}, t) \rangle = 0$ and $\langle \eta_\alpha(\textbf{r}, t) \eta_\beta(\textbf{r}', t) \rangle = \zeta\delta_{\alpha\beta}\delta(\textbf{r} - \textbf{r}')\delta(t - t')$, $\zeta = 2\mu k_BT$. The selfish free energy $F_\alpha$ (zero dimensional) could be calculated by comparing \eqref{15} with \eqref{7},
\begin{align}\label{\theequation}
    F_\alpha &= \frac{1}{2}M_\alpha^2 - \frac{1}{2\Tilde{J}}[\ln{(\cosh{(\Tilde{J}M_\alpha + \Tilde{K}_{\alpha\beta}M_\beta + \Tilde{h}}})) \nonumber\\
    &\quad+\ln{(\cosh{(\Tilde{J}M_\alpha + \Tilde{K}_{\alpha\beta}M_\beta - \Tilde{h}))}}].
\end{align}
Equations \eqref{7} are numerically solved and substituted in \eqref{16} to plot the dynamical free energy.  Fig.~\ref{12}(a) illustrates how droplet formation initiates within the system, giving rise to droplet-induced switching among a set of discrete metastable states. These localised excitations act as triggers that destabilise static order and drive the system through cyclic transitions between eight distinct configurations (clockwise in the MFT phase portrait shown in Fig.~\ref{12}(b)). The presence of this eight-state cyclic behaviour, predicted by MFT (using \eqref{15} and \eqref{16}), is confirmed through 3D Monte Carlo simulations (see Fig.~\ref{12}(d)).

\section{Summary}
To summarise, both MFT and EFT successfully capture the qualitative structure of the bifurcation diagram, including the transition from a Hopf bifurcation to an SNLC bifurcation, separated by a nonequilibrium tricritical (Bautin) point. However, they differ quantitatively from the results obtained via 3D kinetic Monte Carlo simulations. In particular, the presence of all the spin correlations and disorder under competing kinetics enhances the tendency toward discontinuous behaviour and shifts the tricritical point to higher values of $\tilde h$, beyond those predicted by MFT or EFT. Furthermore, EFT predicts a narrower first-order region than MFT. All three approaches agree on a key feature of the phase diagram: above the tricritical point, there exists a threshold value of nonreciprocity, $\tilde K_c$, below which the conventional swap phase disappears. In this regime ($\tilde h > \tilde h_c$ and $\tilde K < \tilde K_c$), the system instead enters a droplet-induced swap phase, cycling among either eight or four metastable states. The cyclic switching between eight metastable states is a new phenomenon, which is not observed in the zero-field nonreciprocal Ising model \cite{avni2025dynamical}. The droplet-induced swap behaviour can be interpreted in terms of the system's dynamical free-energy landscape, where fluctuations enable transitions between local minima. The existence of this regime is confirmed by simulations, which consistently show its emergence across a broad range of parameters.

Several important questions remain open. Most notably, the precise bounds of the interval containing the nonequilibrium tricritical point—and the determination of its universality class through the extraction of critical exponents—are still to be addressed. A systematic exploration of how both the tricritical location and the associated scaling behaviour depend on the width and tail properties of the disorder is also needed, particularly by examining the double-Gaussian random-field distribution over a range of standard deviations $\sigma$. Moreover, the long-term behaviour and scaling characteristics of the droplet-induced swap regime—especially in three dimensions and as a function of the nonreciprocity parameter $\tilde{K}$—remain unclear. Resolving these issues will require extensive Monte Carlo simulations supported by thorough finite-size scaling, Binder-cumulant analyses, and histogram-based diagnostics.

\appendix

\section{Calculation of the coefficients $A_p^\alpha$} \label{Appendix_A}
The coefficients $A_p^\alpha$ are given by
\begin{align}
    A_p^\alpha &= \frac{6!}{p!(6-p)!}\cosh^{6-p}(\tilde J D_x)\sinh^{p}(\tilde J D_x)\nonumber\\
    &\times\big[\cosh(\tilde K_{\alpha\beta} D_x) + M_\beta\sinh(\tilde K_{\alpha\beta} D_x)\big]\,F(x)\big|_{x=0},
\end{align}
where $D_x=\mathrm{d}/\mathrm{d}x$ and $F(x)=\tfrac{1}{2}\big[\tanh(x+\tilde h)+\tanh(x-\tilde h)\big]$.
Below we show the calculation for a representative coefficient and then quote the results for the remaining coefficients.

For $p=0$ we obtain
\begin{align}
    A_0^\alpha &= \cosh^6(\tilde J D_x)\big[\cosh(\tilde K_{\alpha\beta} D_x) \nonumber\\ &+ M_\beta \sinh(\tilde K_{\alpha\beta} D_x)\big]F(x)\big|_{x=0}.
\end{align}
Writing the hyperbolic functions in exponential form, then expanding by the binomial theorem and using the property $\exp{(aD_x)}g(x) = g(x+a)$ yields
\begin{align}
    A_0^\alpha &= \frac{1}{2^7}\sum_{m=0}^6 \binom{6}{m}\big[ F(\Lambda_m^+) + F(\Lambda_m^-) \nonumber \\ & + M_\beta\big(F(\Lambda_m^+) - F(\Lambda_m^-)\big)\big],
\end{align}
where $\Lambda_m^\pm=(6-2m)\tilde J \pm \tilde K_{\alpha\beta}$. Further simplification gives
\begin{align}
    A_0^\alpha &= \frac{1}{2^6}\left[\sum_{m=0}^2\binom{6}{m}\big(F(\Lambda_m^+) - F(\Lambda_m^-)\big) \right. \nonumber \\ & \left. \quad +  \binom{6}{3}F(\Lambda_3^+)\right]M_\beta.
\end{align}
Similarly, the remaining coefficients can be written compactly. In general one obtains
\begin{align}
    A_p^\alpha &= \frac{6!}{p!(6-p)!2^6}\sum_{m=0}^2\omega_m^{(p)}\big[F(\Lambda_m^+) + F(\Lambda_m^-)\big], \\
    A_q^\alpha &= \frac{6!}{q!(6-q)!2^6}\Bigg[\sum_{m=0}^2\omega_m^{(q)}\big[F(\Lambda_m^+) - F(\Lambda_m^-)\big] \nonumber \\& + \omega_3^{(q)}F(\Lambda_3^+)\Bigg]M_\beta,
\end{align}
where $p$ is even and $q$ is odd. The coefficient weights $\omega_m^{(r)}$ are given by
\[
\begin{aligned}
\omega_m^{(0)} &= \binom{6}{m}, \\
\omega^{(1)}  &= \{1,\,4,\,5,\,0,\,-5,\,-4,\,-1\},\\
\omega^{(2)}  &= \{1,\,2,\,-1,\,-4,\,-1,\,2,\,1\},\\
\omega^{(3)}  &= \{1,\,0,\,-3,\,0,\,3,\,0,\,-1\},\\
\omega^{(4)}  &= \{1,\,-2,\,-1,\,4,\,-1,\,-2,\,1\},\\
\omega^{(5)}  &= \{1,\,-4,\,5,\,0,\,-5,\,4,\,-1\},\\
\omega_m^{(6)} &= \binom{6}{m}.
\end{aligned}
\]
\section{MFT and EFT phase portraits} \label{Appendix_B}
Phase portraits were generated using \textit{Mathematica} by integrating the MFT/EFT ODEs with \texttt{NDSolve}, plotting solution trajectories with \texttt{ParametricPlot}, displaying the vector field with \texttt{StreamPlot}, and overlaying nullclines from \texttt{ContourPlot}. Parameter scans were performed with \texttt{Manipulate}.

To determine the critical non-reciprocity $\tilde K_c(\tilde h)$, we numerically integrated the mean-field equations on a high-resolution discretised $(\tilde J, \tilde K)$ grid. The rotational order parameter $\bar{S}$ was computed by time-averaging the trajectory after discarding the initial $90\%$ to ensure the system had reached a steady state. We defined $\tilde K_c$ as the minimal $\tilde K$ value for which $\bar{S} \ge 10^{-5}$ for any given $\tilde J$. A distinct advantage of the mean-field approach is the absence of finite-size fluctuations or local droplet nucleation; consequently, a non-zero $\bar{S}$ strictly indicates the macroscopic swap phase. The threshold of $10^{-5}$ was introduced solely to filter out numerical noise arising from finite integration precision, ensuring that the detected phase transition is physical rather than an artefact of the solver.
\begin{acknowledgements}
The authors acknowledge financial support from the Department of Atomic Energy, India, through the Project (RIN4001-SPS).
\end{acknowledgements}

% If you have acknowledgments, this puts in the proper section head.
%\begin{acknowledgments}
% put your acknowledgments here.
%\end{acknowledgments}
% Create the reference section using BibTeX:
\bibliography{KRFNIM.bib}
\end{document}